\documentstyle[aps]{revtex}

\begin{document}
\title{Peculiarity of the Coulombic Criticality ?}
\author{N. V. Brilliantov$^{1,2}$, J. P. Valleau$^{1}$, C. Bagnuls$^{3}$ and C.
Bervillier$^{4}$}
\address{$^{1}$Department of Chemistry, University of \\
Toronto, Toronto, Canada M5S 3H6\\
$^{2}$ Moscow State University, Physics Department, \\
Moscow 119899, Russia\\
$^{3}$ Service de Physique de l'Etat Condens\'{e}\\
C.E. Saclay, F91191 Gif-sur-Yvette cedex, France\\
$^{4}$ Service de Physique Th\'{e}orique\\
C.E. Saclay, F91191 Gif-sur-Yvette cedex, France}
\maketitle

\begin{abstract}
We indicate that Coulombic systems could correspond to Wilson effective
Hamiltonians similar to that of the ordinary (nonionic) fluids but with a
negative $\varphi ^{4}$-coefficient. In that case, solving the ``exact''
renormalization group equation in the local potential approximation, we show
that close initial Hamiltonians may lead either to a first order transition
or to an Ising-like critical behavior, the partition being formed by the
tri-critical surface. Hence the theoretical wavering encountered in the
literature concerning the nature of the Coulombic criticality may not appear
senseless.

\end{abstract}

In a number of experimental, theoretical and computer simulation studies,
the problem of Coulombic criticality has been addressed \cite{3096}. For
some ionic fluids mean field critical behavior is observed experimentally.
Other systems demonstrate Ising-type behavior, while still other ionic
fluids exhibit a crossover from classical behavior to Ising-like as the
critical point is approached. At the theoretical level, the confused state
of the subject has been well described and clarified by Fisher and Stell 
\cite{3096}.

Coulombic criticality may be studied within the restricted primitive model
(RPM: equal numbers $N_{+}$ and $N_{-}$ of positive and negative hard
spheres of equal diameter, $d$, with charges, $\pm ze$, immersed in a
structureless solvent of dielectric permittivity $\epsilon $; $z/\epsilon=1$
in what follows). If one generally agrees on the existence of a
liquid-gas-like transition at low concentration and low temperature in RPM,
it is still not evident what kind of critical behavior the model actually
possesses. Mean-field-like or Ising-like critical behavior, crossover from
mean-field to Ising behavior, tricriticality or first order transition: all
these conclusions have been expressed and discussed \cite{3096}.

Our aim, in this letter, is first to compute the Wilson effective
Hamiltonian corresponding to RPM at its critical point and then to apply the
renormalization group (RG) techniques to study it. Let us summarize the main
lines of the calculation of the effective Hamiltonian: the details will be
published elsewhere \cite{brilval}.

The Hamiltonian of the RPM (in units of $k_BT$) is a sum of the repulsive
(hard-core) part and of the coulombic part:

\begin{equation}
H=\frac{\beta }{2}\sum_{i,j}{}^{\prime }\phi _{hc}\left( r_{ij}\right) +%
\frac{1}{2}\sum_{\vec{k}}{}^{\prime }\nu (k)\left( \rho _{\vec{k}}\rho _{-%
\vec{k}}-\rho \right)  \label{totHam}
\end{equation}
here $\phi _{hc}\left( r_{ij}\right) $ are the hard-core interactions, $%
\left\{ \vec{r}_{i}\right\} $ denote coordinates of the $N=N_{+}+N_{-}$
particles, $\vec{r}_{ij}=\vec{r}_{i}-\vec{r}_{j}$, and $\rho =N/\Omega $ is
the density, where $\Omega $ is the volume of the system. The coulombic
interactions in Eq.(\ref{totHam}) are written in terms of the {\it charge
density } fluctuation amplitudes, $\rho _{\vec{k}}=n_{\vec{k}}^{a}-n_{\vec{k}%
}^{b}$. The latter are expressed in terms of the amplitudes of {\it density}
fluctuations of positive ($a$), $n_{\vec{k}}^{a}=\Omega ^{-\frac{1}{2}%
}\sum_{j=1}^{N/2}\exp \left\{ -i\vec{k}\vec{r}_{j}^{a}\right\} $, and
negative ($b$), $n_{\vec{k}}^{b}=\Omega ^{-\frac{1}{2}}\sum_{j=1}^{N/2}\exp
\left\{ -i\vec{k}\vec{r}_{j}^{b}\right\} $, particles. In Eq.(\ref{totHam}) $%
\nu (k)\equiv \left( 4\pi e^{2}/k_{B}T\right) /k^{2}$, and the primes over
sums denote that terms with $i=j$ in the first sum and with $\vec{k}=0$ in
the second sum are excluded.

Using the Hubbard-Schofield scheme \cite{hubbard}, we map the ionic fluid
Hamiltonian into a Wilson effective Hamiltonian similar to that obtained for
ordinary (nonionic) fluids \cite{hubbard} which, discarding the derivatives
of the field (local potential approximation), has the following form:

\begin{equation}
H=\int d\vec{r}\left[ \frac{1}{2}\left( \nabla \varphi (\vec{r})\right)
^{2}+V\left( \varphi (\vec{r})\right) \right]  \label{efh2}
\end{equation}
with the ``potential function'':

\begin{equation}
V\left( \varphi \right) =\frac{1}{9\pi ^{2}}\sum_{n=1}^{\infty }\frac{b^{2n}%
}{(2n)!}u_{2n}\varphi ^{2n}  \label{RPM}
\end{equation}
where $b^{2}=4\pi \left( 3\pi \right) ^{2/3}\rho ^{*1/3}/T^{*}$ ($\rho
^{*}=\rho d^{3}$ is the reduced density and $T^{*}=k_{B}Td^{3}/e^{2}$ is the
reduced temperature). As for the ordinary fluid ~\cite{hubbard}, the
coefficients $u_{2n}$ may be expressed in terms of the cumulant averages, $%
\left\langle \rho _{\vec{k}_{1}}\ldots \rho _{\vec{k}_{n}}\right\rangle
_{cR} $, where averaging is performed over the reference system having only
hard--core interactions. First we calculate these quantities for the {\it %
lattice--gas} model and find that $u_{2n}=(-1)^{n+1}$ which yield the usual
Sine-Gordon Hamiltonian for the RPM ~\cite{klein}. To perform evaluation of
the {\it off--lattice} coefficients $u_{2n}$ we use a symmetry of the RPM
with respect to the hard-core interactions, the definitions for the
correlation functions, ~\cite{barker,gray}, and express the coefficients in
terms of the Fourier transforms (taken at zero wave--vectors) of the
``cluster'' functions of the reference hard--sphere system. In obvious
notations ~\cite{gray} these read:~ $h_{2}(1,2)\equiv g_{2}(1,2)-1$, $%
h_{3}(1,2,3)\equiv g_{3}(1,2,3)-g_{2}(1,2)-g_{2}(1,3)-g_{2}(2,3)+2$, etc.,
where $g_{n}(1,\ldots ,n)$ are $n$-particle correlation functions. Using the
relation between the $g_{n+1}$ and $g_{n}$ ~\cite{gray},

\begin{equation}
\chi \rho^2 \frac{ \partial}{\partial \rho} \rho^n g_n = \beta \rho^l \left[
n\, g_n+\rho \int d \vec{r}_{n+1} \left( g_{n+1}-g_{n} \right) \right],
\end{equation}
where $\chi =\rho ^{-1}\partial \rho /\partial P$ is the compressibility, we
iteratively express the Fourier transforms of $h_{n}$ at zero wave--vectors, 
$\tilde{h}_{n}(\vec{0})$, in terms of $\tilde{h}_{n-1}(\vec{0})$ and its
density derivative, and ultimately in terms of $\tilde{h}_{2}(0)$ and its
density derivatives. Then we use the relation ~\cite{gray} $\rho \tilde{h}%
_{2}(0)=\rho k_{B}T\chi -1\equiv z_{0}$, and obtain coefficients for the 
{\it off--lattice } effective Hamiltonian. In particular, $u_{2}=1$, $%
u_{4}=-\,(1+3z_{0})$, $u_{6}=1+15(z_{0}^{2}+z_{0}z_{1}+z_{1})$, $\ldots $
where $z_{1}\equiv \rho \left( \partial z_{0}/\rho \right) ,\ldots $ To
obtain $z_{0}$ one can use the virial expansion for the hard-sphere
pressure, $P/\rho k_{B}T=1+\sum_{k}B_{k}\rho ^{k}$ with the coefficients $%
B_{1},\ldots B_{6}$ known ~\cite{barker} (for small densities), or the
Carnahan-Starling equation of state ~\cite{barker,gray}. Therefore, applying
the above scheme, {\it all} the coefficients of the effective potential $%
V\left( \varphi \right) $ may in principle be found.

We have studied the density dependence of $u_{2n}$ up to $2n=14$ and
observed that all the coefficients are {\it negative} in the density
interval $\sim 0.07\sim 0.09$ where the critical density of the RPM is
expected to be. We have then performed an {\it empirical} analysis and found
that the boundaries of the density interval where the coefficients $u_{2n}$
are negative depend fairly linearly on $1/n$ (see Fig.~1). Extrapolating
this dependence we have found that all the coefficients become {\it positive 
} for $n>22$ (see Fig.~1). Being secure in the knowledge that the effective
Hamiltonian is bounded from below, we can envisage a RG analysis.

We thus consider the ``exact'' RG equation in the local potential
approximation~\cite{hashas} in three dimensions which, using the same
notations as in \cite{3554}, reads: 
\begin{equation}
\dot{f}=\frac{1}{4\pi ^{2}}\frac{f^{\prime \prime }}{1+f^{\prime }}-\frac{1}{%
2}yf^{\prime }+\frac{5}{2}f  \label{4}
\end{equation}
in which $y$ stands for the dimensionless field and $f(y,l)=\partial
V(y,l)/\partial y$, $f^{\prime }=\partial f/\partial y$, $f^{\prime \prime
}=\partial ^{2}f/\partial y^{2}$, $\dot{f}=\partial f/\partial l$ with $l$
the RG scale parameter (that relates two different ``momentum'' scales of
reference such that $\Lambda _{l}=e^{-l}\Lambda _{0}$).

Ideally, the question raised may be formulated as follows: considering the
effective Hamiltonian for the RPM at its assumed critical point (taken e.g.
from Monte Carlo data \cite{valtor,cail}) as an initial Hamiltonian ($l=0$)
for Eq. (\ref{4}), will the solution of Eq. (\ref{4}) flow toward the Ising
fixed point [the unique non-trivial fixed point of Eq.(\ref{4})] or not?

Unfortunately, considering the function $f(l=0,y)$ which gives the initial
conditions of RPM for Eq.(\ref{4}) (at $\rho_c^*=0.0857$, $T_c^*=0.052$ ~%
\cite{valtor}, or $\rho_c^*=0.080$, $T_c^*=0.0488$ ~\cite{cail}), we have
observed that the denominator $1+f^{\prime }(y)$ in (\ref{4}) has
singularities in the interval, $0.1<y<0.12 $. Moreover, Eq.(\ref{4}) does
not allow us to handle values of Hamiltonian coefficients as large as those
of Eq.(\ref{RPM}). Considering that Eq. (\ref{4}) is an approximation, that
impossibility has no particular significance relative to the behavior of RPM.

However, we have considered that the negative value of $u_{4}$ found for RPM
could be a characteristic feature of qualitative importance for (some) ionic
systems. Thus we turn our attention to the solutions of Eq.(\ref{4}) with
initial functions involving negative values of $u_{4}$.

In \cite{3554} a detailed study of the approach to the Ising fixed point
using Eq. (\ref{4}) has been presented. However all the initial Hamiltonians
considered were taken with $u_{4}>0$. To our knowledge, the few studies
based on RG techniques that have, up to now, considered negative values of $%
u_{4}$, either perturbatively \cite{perturb} in three dimensions or, more
recently, non-perturbatively \cite{3759} in four dimensions, have concluded
that there is no stable fixed point (the Ising fixed point cannot be reached
starting with $u_{4}<0$) and thus to the lack of ``true'' criticality (there
would be no divergent correlation length, like in a first order transition).

To be short, we consider the following simple functions as initial
conditions to Eq. (\ref{4}) (a detailed study of the case $u_{4}<0$ will be
published elsewhere \cite{bb}):

\begin{equation}
f(y,0)=u_{2}(0)y+u_{4}(0)y^{3}+u_{6}(0)y^{5}  \label{fexp}
\end{equation}
corresponding to a point of coordinates $%
(u_{2}(0),~u_{4}(0),~u_{6}(0)~,0~,0,~\cdots )$ in the space ${\cal S}$ of
Hamiltonian coefficients (the dimension of ${\cal S}$ is infinite). Since we
want to set $u_{4}(0)<0$, and at least one positive higher term is needed to
make the Hamiltonian bounded from below, we choose to set $u_{6}(0)$
positive. Having chosen a (negative) value for $u_{4}(0)$ and a (positive)
value for $u_{6}(0)$, we use the ``shooting'' method\thinspace \cite{3554}
to determine the critical value $u_{2}^{c}(0)$ of $u_{2}(0)$ which brings $%
f(y,0)$ in the critical subspace ${\cal S}_{\text{c}}$ of ${\cal S}$. The
``shooting'' method is based on the fact that, for sufficiently large values
of $l$, the RG trajectories go away from ${\cal S}_{\text{c}}$ in two
opposite directions according to the sign of $u_{2}(0)-u_{2}^{c}(0)$.

Let us summarize our results by considering the case $u_{4}(0)=-6$ as an
example (see Fig. 2).

\begin{description}
\item[A]  If $u_{6}(0)=16$, we find $0.3836174>u_{2}^{c}(0)>0.3836151.$ The
associated RG trajectory goes away from the Gaussian fixed point $P_{\text{G}%
}$ and remains in the sector $u_{4}<0$ of ${\cal S}_{\text{c}}$. Hence it
never reaches the Ising fixed point that lies in the sector $u_{4}>0$ .
Instead the trajectory is attracted to a stable submanifold of dimension one
(an infra-red stable trajectory) that emerges from $P_{\text{G}}$. On a pure
field theoretical point of view, this trajectory is a {\em renormalized}
trajectory (let us denote it by $T_{u_{4}<0}$) to which is associated the
continuum limit of an asymptotically free scalar field theory in three
dimensions. It is very likely that $T_{u_{4}<0}$ is the continuation below
four dimensions of the continuum limit recently studied on a lattice in \cite
{3759} and corresponding to a scalar theory with negative quartic
interaction. {\em The lack of any fixed point ending }$T_{u_{4}<0}${\em \
means that the correlation length remains finite at the assumed critical
point}. This situation could be compared to the fact that no heat--capacity
divergency has been observed in the Monte Carlo study of the RPM ~\cite
{valtor}.

\item[B]  If $u_{6}(0)=20$, we find $0.30131>u_{2}^{c}(0)>0.30122$ and the
associated RG trajectory goes toward the (Wilson--Fisher) Ising fixed point
and approaches it along the usual renormalized trajectory associated with
the continuum limit of the scalar field theory in three dimension usually
called the $\varphi _{3}^{4}$-field theory. This renormalized trajectory
(denoted by $T_{1}$ in \cite{3554}) interpolates between the Gaussian and
the Ising fixed points. Hence {\em there exist initial Hamiltonians with }$%
u_{4}<0${\em \ that belong, nevertheless, to the basin of attraction of the
Ising-like fixed point}.

\item[C]  Between the two preceding cases, we find a trajectory (with $%
u_{6}(0)=18.3125\cdots $ and $0.3324573>u_{2}^{c}(0)>0.3324549$) that
directly flows towards $P_{\text{G}}$ (it is neither attracted to $%
T_{u_{4}<0}$ nor to $T_{1}$). That kind of initial Hamiltonian obtained by
adjusting two coefficients ($u_{2}(0)$ and $u_{6}(0)$) lies on the {\em %
tri-critical subspace} ${\cal S}_{\text{t}}$ of ${\cal S}$. Any trajectory
on ${\cal S}_{\text{t}}$ approaches $P_{\text{G}}$ along a unique
(attractive) trajectory that imposes the required very slow (logarithmic)
flow in the vicinity of $P_{\text{G}}$ (see Fig. 2).
\end{description}

So, it appears that for $u_{4}(0)<0$, very close Hamiltonians may lead to
very different behaviors and this feature is due to the vicinity of the
tricritical subspace. If one adopts the idea that effective Hamiltonians for
Coulombic systems may be characterized by a negative value of $u_{4}$, then
it is not amazing that the tricritical state has been the center of a recent
discussion relative to the nature of the Coulombic criticality. Also, it is
easy to verify that the experimental observations of mean field values for
the critical indices might be due to a retarded crossover towards the Ising
behavior ~\cite{fish}. One may observe on Fig. 2 that the trajectory
corresponding to case B follows a long path to the Ising fixed point (and
passes close to $P_{\text{G}}$) compared to a trajectory that could
correspond to a simple fluid. In order to illustrate qualitatively that
retarded crossover, we have drawn in Fig. 3 the evolutions [calculated from
Eq. (\ref{4})] of a pseudo effective exponent $\nu _{\text{eff}}\left( \tau
=\left| u_{2}(0)-u_{2}^{c}\right| /|u_{2}^{c}|\right) $associated to two
effective Hamiltonians with $u_{4}(0)=-6$ and $u_{6}(0)=20$ on one hand
[RPM-like ? (B)] and $u_{4}(0)=+3$ and $u_{6}(0)=0$ on the other hand
(Simple fluid).

Helpful discussions with H. Stoof, G. Stell and B. Lee are gratefully
acknowledged. The work was partly supported by Natural Science and
Engineering Research Council of Canada.

\begin{center}
FIGURE CAPTIONS
\end{center}

\begin{description}
\item[Figure 1]  Dependence of the boundaries of the density interval where
the coefficients $u_{2n}$ of the effective Hamiltonian are negative as a
function of $1/n$. Extrapolation suggests that {\it all} the coefficients
with $n>22$ are positive. (Note that the density at which the ``negative''
interval shrinks to zero, $\rho =0.0856$ is very close to the critical
density from the Monte Carlo data ~\cite{valtor,cail}).

\item[Figure 2]  Projection onto the plane $\{u_{2},u_{4}\}$ of various RG
trajectories (in ${\cal S}_{\text{c}}$) obtained by solving Eq. (\ref{4}).
Black circles represent the Gaussian ($P_{\text{G}}$) and Ising (IFP) fixed
points. The ideal trajectory (dot line) which interpolates between these two
fixed points represents the renormalized trajectory (RT) of the so-called $%
\phi _{3}^{4}$ field theory in three dimensions (usual RT). White circles
represent the projections onto the plane of initial (unrenormalized)
critical Hamiltonians. For $u_{4}(0)>0$, some effective Hamiltonians run
toward the Ising fixed point asymptotically along the usual RT (simple
fluid). Instead, for $u_{4}(0)<0$ and according to the values of Hamiltonian
coefficients of higher order, the RG trajectories either (A) meet an endless
RT emerging from $P_{\text{G}}$ (dashed curve) and lying entirely in the
sector $u_{4}<0$ or (B) meet the usual RT towards IFP. The frontier which
separates these two very different cases (A and B) corresponds to initial
Hamiltonians lying on the tri-critical subspace (white square C) that are
source of RG trajectories flowing toward the $P_{\text{G}}$ asymptotically
along the tricritical RT. Notice that the coincidence of the initial point B
with the RG trajectory starting at point A is not real (it is accidentally
due to the projection onto a plane). The restricted primitive model could
correspond either to the case A or to the case B.

\item[Figure 3]  Sketchy representation of the effective exponents $\nu _{%
\text{eff}}(\tau )$ calculated in the local potential approximation along
the two RG trajectories which, on Fig. 2, both flow towards the Ising fixed
point [``RPM-like ? (B)]'' corresponds to $u_{4}(0)<0$ and ``Simple fluid''
to $u_{4}(0)>0$). The full lines roughly indicate the experimentally
accessible parts in each case.
\end{description}


\begin{references}
\bibitem{3096}  Experimental reviews: K. S. Pitzer, Acc. Chem. Res. {\bf 23}%
, 333 (1990); J. M. H. Levelt-Sengers and J. A. Given, Molecular Phys. {\bf %
80}, 899 (1993); H. Weing\"{a}rtner, M. Kleemeier, S. Wiegand and W.
Schr\"{o}er, J. Stat. Phys. {\bf 78}, 169 (1995); H. Weing\"{a}rtner and W.
Schr\"{o}er, J. Mol. Liq. {\bf 65-66}, 107 (1995). Theoretical reviews: M.
E.~Fisher, J. Sta. Phys. {\bf 75}, 1 (1994); J. Phys.: Cond. Matt., {\bf 8},
9103 (1996); G.~Stell, J. Phys.: Cond. Matt., {\bf 8}, 9329 (1996);
J.Stat.Phys., {\bf 78}, 197 (1995).

\bibitem{brilval}  N. V.~Brilliantov, J. P.Valleau, in preparation.

\bibitem{hubbard}  J.~Hubbard and P.~Schofield, Phys. Lett., {\bf A40}, 245
(1972).

\bibitem{klein}  H.~Kleinert, {\it Gauge Fields in Condensed Matter}, World
Sci., Singapore, 1989, v.1, Chap.7. Some (unimportant) difference of our
result occurs because the present calculations are performed for the
canonical ensemble.

\bibitem{barker}  J. A.Barker and D. Henderson, Rev. Mod. Phys., {\bf 48},
587 (1976).

\bibitem{gray}  C. G.Gray and K. E.Gubbins, {\it Theory of molecular fluids}%
, Clarendon Press, Oxford, 1984.

\bibitem{hashas}  A. Hasenfratz and P. Hasenfratz, Nucl.\ Phys. {\bf B270
[FS16]}, 687 (1986).

\bibitem{3554}  C. Bagnuls and C. Bervillier, J. Phys. Stud. {\bf 1}, 366
(1997); see also Phys. Rev. {\bf B41}, 402 (1990).

\bibitem{valtor}  J. P.Valleau and G. M.Torrie, in preparation.

\bibitem{cail}  J. M.Caillol, D. Levesque and J. J. Weis, J. Chem. Phys., to
appear

\bibitem{perturb}  E. K.Riedel and F. J. Wegner, Phys. Rev. Lett., {\bf 29},
349 (1972); Phys. Rev. {\bf B9}, 294 (1974); A. I.Sokolov, Sov. Phys. JETP, 
{\bf 50}, 802 (1979).

\bibitem{3759}  K. Langfeld and H. Reinhardt, hep-ph/9702271, submitted to
Phys. Lett. B.

\bibitem{bb}  C. Bagnuls and C. Bervillier, in preparation.

\bibitem{fish}  See M.\ E.\ Fisher in \cite{3096} and references therein.
\end{references}
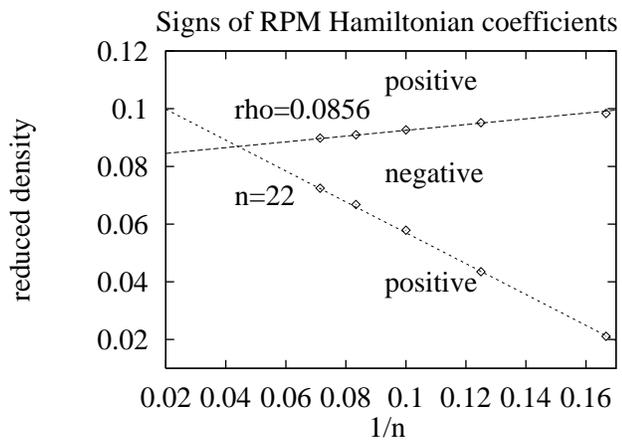
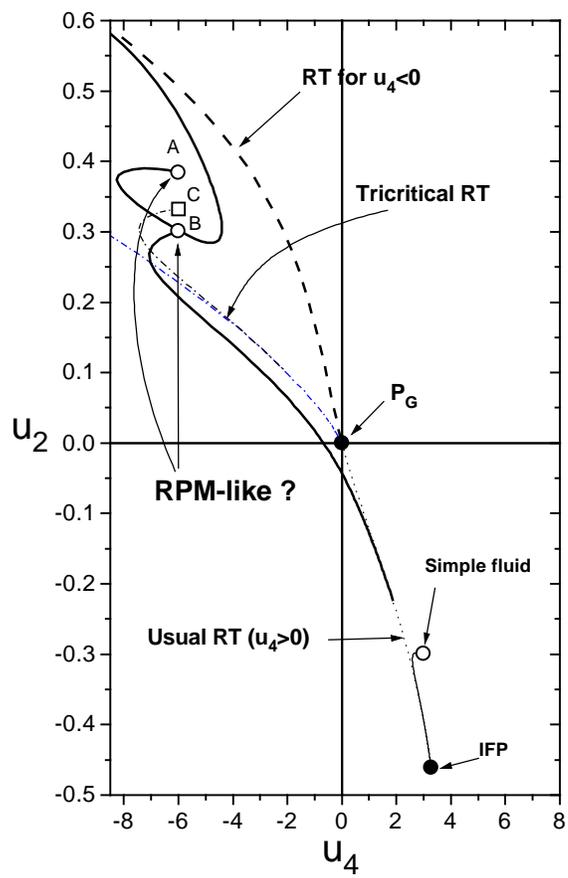
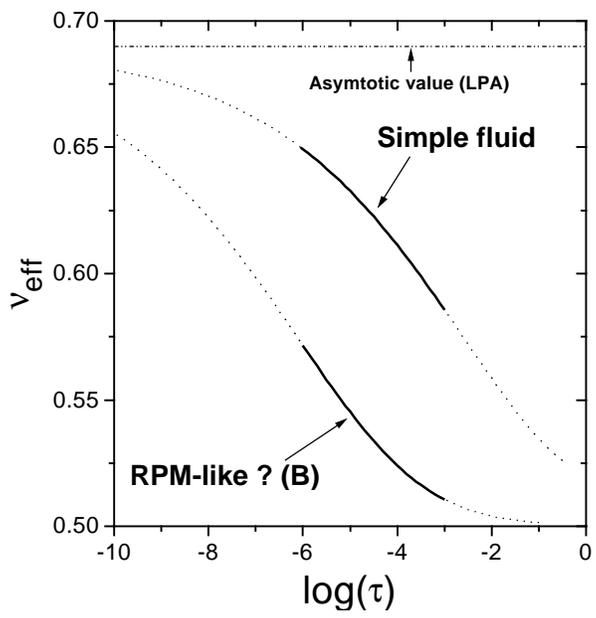
\end{document}